\documentclass[preprint,superscriptaddress,aps,showpacs,floatfix]{revtex4-1}
\usepackage{amsmath,amssymb,amscd}
\usepackage{amsthm}
\usepackage{mathrsfs}
\usepackage{bbm}
\usepackage{bm}
\usepackage{cancel}
\usepackage{graphicx}
\usepackage{multirow}
\usepackage{array}
\usepackage{courier} 
\usepackage{afterpage}
\usepackage{ifthen}
\usepackage{fancyhdr}
\usepackage{color}
\usepackage{textcomp}
\usepackage[bookmarks,plainpages=false]{hyperref}
\begin{document}
\title{ Fresh study of simultaneous electron-photon excitation of a Hydrogen atom based on Bethe-Born approximation}
 \author{Behnam Nikoobakht}
  \affiliation{Theoretische Chemie, Physikalisch-Chemisches Institut, Universit\"{a}t Heidelberg, INF 229, D-69120 Heidelberg, Germany}

\begin{abstract}
The advent of powerful laser sources has made it possible to observe a relatively large cross section of 
the excited state of Hydrogen atom. This is due to the effect of joint collisions of a linearly polarized 
$N$-photon and high-energy electron. For such a process, we evaluate the excitation cross section for 
geometries, in which the laser field is perpendicular or parallel to the initial momentum of the electron.  
The second-order, time-dependent perturbation theory together with Bethe-Born approximation suitable for
an electron with a large incident energy is employed to obtain the transition amplitude. The amplitudes are 
calculated for the $S-S$ and $S-D$ transitions of the Hydrogen atom in the Sturmian representation of the 
non-relativistic Green's function. In particular, we investigate the excitation cross sections for 
transitions, which have an initial state $1S$ and final state $nS$ with $n\in \lbrace 2,3,4,5\rbrace$. 
The characteristic dependence of the excitation cross section on the momentum of the projectile is shown 
and discussed. Our investigation indicates that the Bethe-Born approximation yields reasonable results for the
excitation cross section of the simultaneous electron photon excitation process when a high energy projectile
is treated. 
\end{abstract}
\maketitle
\section{Introduction}
The influence of an external magnetic field on atoms was investigated in the Thompson or Rayleigh
scattering formulated in the framework of classical electrodynamics~\cite{V1962,Ig1964,Je1969}. 
This was the first attempt to show the significant effect of the photon field on atomic collisions. 
This classical result has been confirmed by quantum mechanical treatments, which are also paved 
the way for the creation of a new class of scattering process involving simultaneous collision of 
three partners, electron, photon and atom~\cite{Mm1993}. This collision process belongs to the
category of the non-collective processes, in which only three particles are involved in the 
scattering~\cite{Mm1993,FhAjJk1998}. The main advantage of this scattering process is to provide a 
theoretical and experimental base for the study of collective processes including more than three 
particles interacting simultaneously, such as plasma~\cite{Nm1993}. Performing a thorough investigation 
on the non-collective process is required to gain more insight into the collective processes, such as 
the laser heating of a plasma and laser-driven fusion~\cite{Sg1977,KbSj1974}.

In this paper, it is our purpose to study one of the non-collective processes, which is a simultaneous 
electron-photon excitation (SEPE). For such a process, the excitation cross sections depend on the energy 
of an incoming electron, the intensity, frequency as well as photon polarization of the laser field~\cite{FhAjJk1998,Km1981,PfYa1988}. This unique behavior leads  to more substantial cross sections 
than that of the usual excitation process of atomic systems, where the interaction of the electron and 
atom is only treated in the absence of the laser field~\cite{Mm1993,FhAjJk1998,Km1981}. 

\begin{figure}[htb]
\begin{center}
\begin{minipage}{0.7\linewidth}
\begin{center}
\includegraphics[width=1\linewidth]{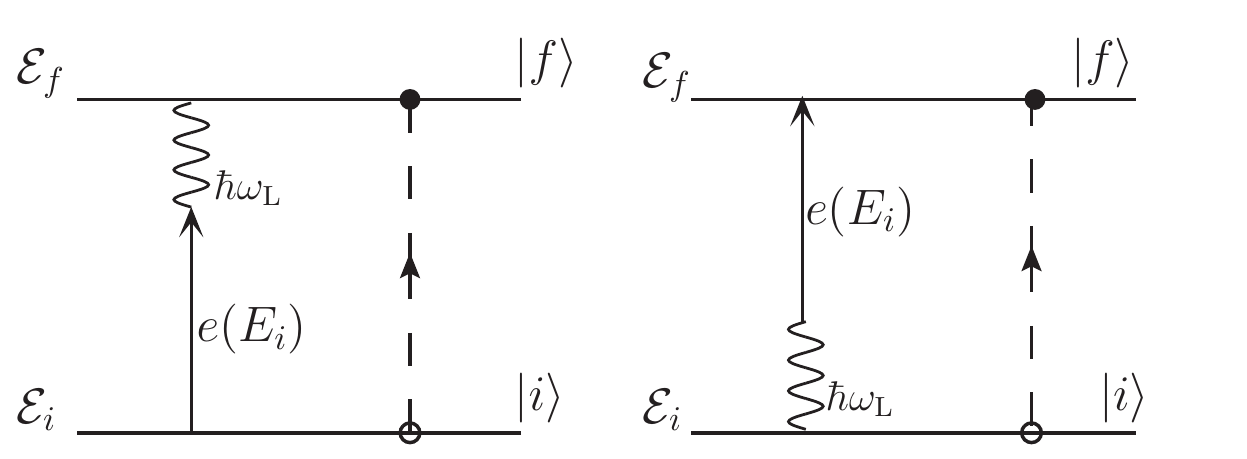}
\caption{\label{fig1}
Schematic representation of the SEPE process. The electron adds the finite 
value of the energy $E_i$ the photon energy of the laser field $\omega_{\rm L}$ making 
possible to take place the transition between initial state $\vert i\rangle$ and final 
state $\vert f \rangle$ in a Hydrogen atom.}
\end{center}
\end{minipage}
\end{center}
\end{figure}

The idea of the SEPE process was innovated by Buimstrov in 1969~\cite{Vb1969}. He proposed that if the 
photon energy of the laser field is not sufficient to excite an atom from its ground 
state to excited state, an additional free electron may add the energy missing, see Fig.~\ref{fig1}. 
With the advent of powerful laser sources, it has become possible to detect the second-order SEPE 
process, which has larger excitation cross section than the first-order process~\cite{Vb1969,NrFf1978,NrFf1976}. 
There have been many attempts in order to calculate the excitation cross section of the second-order SEPE process~\cite{NrFf1978,NrFf1976}. The authors in Ref.~\cite{NrFf1978} considered a high energy electron 
(projectile) colliding with a Hydrogen atom in the presence of the laser filed. They assumed that the 
electron and photon energies coincide with the ground and excited state energy difference of the Hydrogen 
atom [see Eq.~(\ref{resoncon})]. The authors discussed resonance structures of the excitation cross 
section of the SEPE process in the second-order perturbation theory, where the interactions of the 
atom with the electron and photon distort the atomic states. The atom-electron interaction was investigated 
in the framework of the Bethe-Born approximation, where the dipole term of the free electron and the 
bound-atomic electron interaction gives an important contribution. The atom-photon interaction was dealt 
with in the dipole approximation, where it was assumed that the wave length of the laser light is 
large in comparison with the spatial extent of atomic wave functions.

The transition amplitude of the SEPE process presented in Ref.~\cite{NrFf1978} was calculated in 
the velocity gauge. The result was limited to the transitions between states with the same parity ~({\it i.e.}, 
$S$-$S$ transition) for the SEPE process in the Hydrogen atom. In this work, we want to investigate
the SEPE process and to go beyond the results presented in Ref.~\cite{NrFf1978}; we first 
want to obtain analytically the transition amplitude of the SEPE process of the Hydrogen atom for the $S-S$ 
and $S-D$ transitions. Second, we evaluate numerically the total excitation cross sections of the SEPE 
process, in which transitions from an initial state to highly excited states are dealt with  for two 
different geometries in our investigation. In Sec.~\ref{genral}, we reconsider the SEPE process introduced 
in Ref.~\cite{NrFf1978} and compute the transition amplitude in the length gauge. As we expect, the same 
result for the transition amplitude as presented in Ref.~\cite{NrFf1978} is obtained. Going beyond the result of Ref.~\cite{NrFf1978}, our finding for the transition amplitude of the second-order SEPE process can be 
applied for states with different parities $(\Delta l=0,\pm 2)$. Then, we restrict ourselves to transitions 
with $\Delta l=0$ ({\it e.g.} S-S transition) and use the Sturmian representation of the non-relativistic 
Coulomb Green's function to evaluate the total cross section of the SEPE process. The analytical results for 
the radial integral appearing in the evaluation of the total cross sections are presented in Appendix~\ref{appendixA}. 
In Sec.~\ref{neumeri}, the numerical calculations of the excitation cross sections for the $1S-nS$ transitions, $n\in \lbrace 2,3,4,5\rbrace$ induced by the SEPE process for two different geometries are described. In these numerical evaluations, the 
characteristic dependence of the excitation cross section on the energy of the free electron and on the 
scattering angle $\theta$ for two different geometries is presented and discussed. The effect of the laser 
orientations with respect to the initial momentum of the free electron on the excitation cross sections is also 
disputed. We also compare our results with the literature and discuss the reasons for inconsistencies, 
which are illustrated in Appendix~\ref{appendixB}. Summary and conclusion are given in Sec.~\ref{summ}.
\section{General description}
\label{genral}
In this section, we shall give a general description of the excitation of the Hydrogen atom via simultaneous 
collisions of a high-energy electron with an initial momentum $\bm{k}_i$ and a linearly polarized laser field with 
the frequency $\omega_{\rm L}$, 
\begin{align}
e(k_i)+H(n_i)\pm l\hbar \omega_{\rm L}\rightarrow e(k_f)+ H(n_f),
\label{reaction}
\end{align}
where $n_i$ and $n_f$ refer to the principal quantum numbers of the initial and final states in the Hydrogen 
atom. Note that $l$ is the number of emitted or absorbed photons in this process and $k_f$ is the final 
momentum of the free electron.

As described in details in Refs.~\cite{Nm1993,Vb1969}, the Hydrogen atom in a joint electron-photon 
collision is populated in an excited state in the following steps: A free electron with momentum ${\bm k}_i$ 
penetrates the laser field, which is switched on adiabatically (slowly). Then, the electron emits or 
absorbs $l$ photons and its energy reaches $k_i^2/2m\pm l\hbar\omega_{\rm L}$. In the next step, the 
electron interacts with the Hydrogen atom in the initial state and the excitation process takes 
place. The final energy of the free electron reads, 
\begin{align}
E_f=k_i^2/2m-\Delta {\cal E} \pm l\hbar\omega_{\rm L},
\label{resoncon}
\end{align} 
where $\Delta \cal E$ is the energy difference between the final and initial states. At the final 
stage, we turn off adiabatically the laser field and the free electron with energy 
$k_i^2/2m\pm n\hbar\omega_{\rm L}$ leaves the Hydrogen atom by reabsorption of photons.

As indicated in the previous paragraph, we have used the fact that the time-dependent perturbation is gradually 
turned on and off. This is the so-called adiabatic switching of the interaction, which allows to 
avoid the effect of an abrupt change in the Hamiltonian, see Eq.~(\ref{totalhamiltonian})~\cite{Jj1994}. 
This allows to replace the time-dependent potential $V(t)$ in the Hamiltonian by $V^{\rm ad}(t)$,
\begin{align}
V(t)\rightarrow V^{\rm ad}(t)= e^{-\eta \vert t\vert}V(t),
\label{dampp}
\end{align} 
where ${\it ad}$ refers to the adiabatic picture and $\eta$ is the damping parameter, which is in the range $0<\eta<1$. 
In the following section, we shall employ the concept of the adiabatic switching of the interaction in the 
evaluation of the transition amplitude corresponding to the SEPE process of Eq.~(\ref{reaction}).

\subsection{Evaluation of scattering amplitude}
Consider an electron and a plane-wave monochromatic laser field in the $z$ direction colliding simultaneously with 
a Hydrogen atom in the initial state $n_i$,  as described in Eq.~(\ref{reaction}). This three-body inelastic 
scattering is characterized by the following Hamiltonian 
\begin{align}
H&=H_0+V(t),
\label{totalhamiltonian}
\end{align}
where $H_0$ refers to the unperturbed Hamiltonian including free electron (projectile) 
and Hydrogen atom Hamiltonians 
\begin{align}
H_0&=\frac{p_2^2}{2m_e}+\frac{p_1^2}{2m_e}+\frac{Ze^2}{r_1},
\label{unperturbed}
\end{align}
where $Z$ is the nuclear charge. The time-dependent interaction term $V(t)$ in Eq.~(\ref{totalhamiltonian}) 
involves two terms, 
\begin{align}
V(t)&={\cal V}_1+{\cal V}_2.
\label{perturbation}
\end{align}
The first term $ {\cal V}_1$ characterizing the 
interaction of laser field with free and bound electrons is  
\begin{align}
{\cal V}_1&= V_1 \sin(\omega_{\rm L} t) e^{-\eta \vert t \vert}, \nonumber\\
 V_1&=eE{\bm \epsilon} \cdot ({\bm r}_1+{\bm r}_2), 
\label{potentiala}
\end{align}
where $e$, $E$ and ${\bm \epsilon}$ refer to the magnitude of the electric charge, electric field 
and polarization of electric field, respectively. ${\bm r}_1$ and ${\bm r}_2$ denote the position 
of the bound and free electrons, respectively. Note that the concept of the adiabatic switching of the 
time-dependent interaction is considered by the small parameter $\eta$ in Eq.~(\ref{potentiala}) implying  
that the time-dependent perturbation of Eq.~(\ref{potentiala}) vanishes at $t\rightarrow -\infty$ , then 
start to increase very slowly to its full value. At the end of the calculation, we let $\eta \rightarrow 0$ 
to yield the constant intensity result after performing the time integration of the second-order term in the 
Dyson series [see Eq.~(\ref{dyson})]. We now turn our attention to the second term ${\cal V}_2$ of 
Eq.~(\ref{perturbation}). This term refers to the interaction of the free electron and Hydrogen atom and reads 
\begin{align}
{\cal V}_2&=e^2\Big(-\frac{1}{r_2}+\frac{1}{\vert {\bm r}_1-{\bm r}_2 \vert}\Big).
\label{potentialb}
\end{align}
In the interaction picture, where it is convenient to investigate the SEPE process of 
Eq.~(\ref{reaction}), the time-dependent potential is represented by~\cite{Jj1994,Cj1975}  
\begin{align}
{\cal V}^{\rm I}_{1, 2}(t)&=\exp\Big(\frac{i}{\hbar}H_0t\Big) {\cal V}_{1, 2}(t)
\exp\Big(\frac{-i}{\hbar}H_0t\Big)
\label{interpoten}
\end{align}  
The subscript ${\rm I}$ refers to the interaction picture, where the kinematical and dynamical evolution of 
the system are separated. The former one corresponding to the evolution of observable relies on $H_0$ of 
Eq.~(\ref{unperturbed}), while the latter one corresponding to the evolution of the state vectors is governed 
by the time-dependent interaction $V^{\rm I}(t)$ of Eq.~(\ref{perturbation})~ \cite{Cj1975},
\begin{align}
 i\frac{\partial\phi(t)}{\partial t}=V^{\rm I}(t)\phi(t).
\label{evolui}
\end{align}
This relation is called Tomonaga-Schwinger equation. Its solution expressed by evolution operator $U_{\rm I}(\eta,t)$ 
yields Dyson's perturbation expansion~\cite{Cj1975}
\begin{align}
 U_{\rm I}(\eta,t)&=1-i\int_{-\infty}^{\infty}dt_1V^{\rm I}(\eta,t_1)+(-i)^2\int_{-\infty}^{\infty}
dt_1\int_{-\infty}^{t_1}dt_2 T[V^{\rm I}(\eta,t_1)V^{\rm I}(\eta, t_2)]+\cdots,
\label{dyson}
\end{align}
\begin{figure}[htb]
\begin{center}
\begin{minipage}{0.7\linewidth}
\begin{center}
\includegraphics[width=1.1\linewidth]{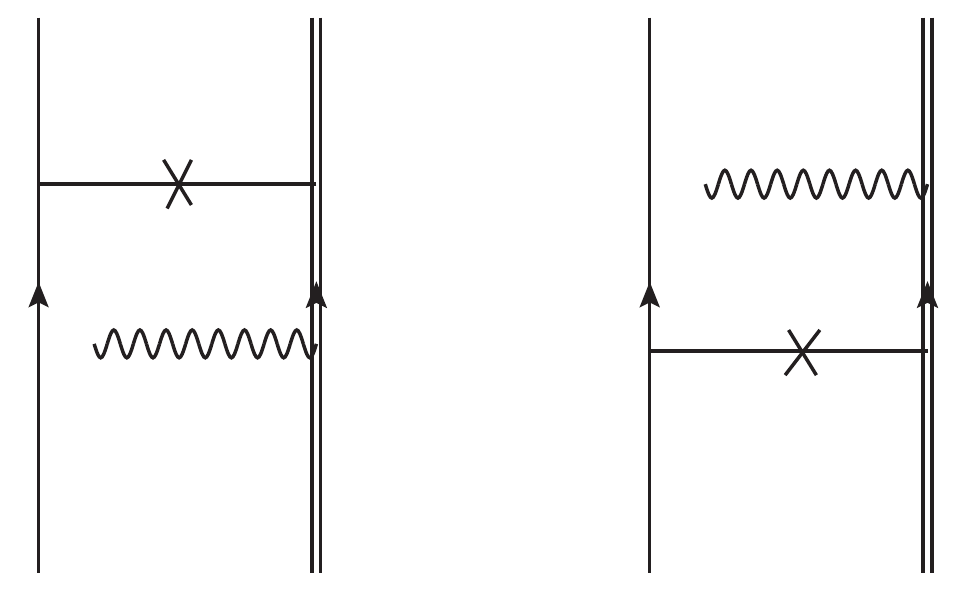}
\caption{\label{feynman}
The corresponding Feynman diagrams to the transition amplitude of 
laser-assisted inelastic scattering of Eq.~(\ref{reaction}). The 
L. H. S diagram is associated with $s_{12}$ and the other one is 
for $s_{21}$ as indicated Eq.~(\ref{secondperturbation}). The single 
and double lines are designated for the free, bound electrons, while the 
wavy line refers to the photon. The line with cross refers to the 
interaction between the free electron and the Hydrogen atom.} 
\end{center}
\end{minipage}
\end{center}
\end{figure}
where $T$ means a chronological ordering operator. Now, we assume that there is an initial time-dependent 
state $\vert \phi^{\rm I}_i(t)\rangle$ in the interaction picture. Based on the assumption of adiabatic switching of 
interaction, the wave function in the remote past is free of the laser field effect, 
{\it i.e.}, $\vert \phi^{\rm I}_i(t\rightarrow -\infty)\rangle=\vert \phi_i\rangle$, where $\vert 
\phi_i\rangle$ is the eigenfunctions of the Hamiltonian of $H_0$ in Eq.~(\ref{unperturbed}). The final 
state wave function in the interaction picture is generated by the evolution operator $U_{\rm I}(\eta,t)$ 
of Eq.~(\ref{dyson}),
\begin{align}
\vert \phi^{\rm I}_f(t)\rangle&=U_{\rm I}(\eta,t)\vert \phi_i\rangle=
\sum_{m} c_m(t)\vert \phi_m\rangle\nonumber\\
 c_m(t)&=\langle m\vert\phi^{\rm I}_f(t)\rangle,
\label{waveexpan}
\end{align}
where $\vert \phi^{\rm I}_f(t)\rangle$ is expanded in a complete set ${\vert \phi_m \rangle}$ of eigenstates of
$H_0$. The corresponding transition amplitude to the reference state, $\vert \phi_f\rangle$, reads
\begin{align} 
 c_{\phi_f}^{if}=\langle \phi_f\vert \phi^{\rm I}_f(t)\rangle=\langle \phi_f\vert 
U_{\rm I}(\eta,t)\vert \phi_i\rangle
\label{transi.amp}
\end{align}
Inserting the evolution operator $U_{\rm I}(\eta,t)$ of Eq.~(\ref{dyson}) into Eq.~(\ref{transi.amp})
and keeping only the second-order perturbation expansion yields 
\begin{align}
c^{if}_{\phi_f}&=\Big(-\frac{i}{\hbar}\Big)^2\Big[\int_{-\infty}^{\infty} dt_1\int_{-\infty}^{t_1}dt_2
\Big\langle \phi_f \Big\vert T[{\cal V}^{\rm I}_1(t_1){\cal V}^{\rm I}_2(t_2)]\Big\vert \phi_i 
\Big\rangle\nonumber\\
&=\Big(-\frac{i}{\hbar}\Big)^2\Big[\underbrace{\int_{-\infty}^{+\infty} dt_1 \int_{-\infty}^{t_1} dt_2
\Big\langle \phi_f \Big\vert {\cal V}^{\rm I}_1(t_1){\cal V}^{\rm I}_2(t_2)\Big\vert \phi_i 
\Big\rangle}_{s_{12}}\nonumber\\
&\quad\mbox{}+\underbrace{\int_{-\infty}^{+\infty} dt_1 \int_{-\infty}^{t_1}
 dt_2 \Big \langle \phi_f \Big \vert {\cal V}^{\rm I}_2(t_1){\cal V}^{\rm I}_1(t_2)
\Big \vert \phi_i \Big \rangle}_{s_{21}}\Big],
\label{secondperturbation}
\end{align}
where the terms $s_{12}$ and $s_{21}$ are associated with the two Feynman diagrams in Fig.~\ref{feynman} 
characterizing the second-order SEPE process of Eq.~(\ref{reaction}). In Eq.~(\ref{secondperturbation}), 
the initial and final eigenfunctions $\vert \phi_{f, i}(r_1,r_2)\rangle$ of the unperturbed Hamiltonian 
$H_0$ composed of the product of the free- and bound state-electron wave functions read
\begin{align}
\vert \phi_{f, i}(r_1,r_2)\rangle&=\vert \psi_{f, i}(r_1)
\rangle\frac{e^{i{\bm k}_{f, i}\cdot {\bm r}_2}}{(2\pi)^{3/2}}.
\label{wavefunction}
\end{align}
Here, $\psi_{f,i}(r_1)$ refer to the bound-electron wave functions in the initial and final (excited) states
and $k_{f,i}$ refers to the momenta of free electron before and after collision~[see Eq.~(\ref{reaction})].
Note that the eigenenergy of the Hamiltonian $H_0$ of Eq.~(\ref{unperturbed}) is
\begin{align}
E_{f,i}&= {\cal E}_{f,i}+\frac{\hbar^2k_{f,i}^2}{2m_e},
\label{eigenenergy}
\end{align}
where the first term  denotes the bound electron energy, while the second one refers to the kinetic
energy of the incoming and outgoing electrons.

In the following, we concentrate on the evaluation of $s_{21}$ in Eq.~(\ref{secondperturbation}) corresponding 
to the R. H. S. diagram in Fig.\ref{feynman}. A similar computational method is applied for the evaluation of
$s_{12}$ in Eq.~(\ref{secondperturbation}), which corresponds to the L. H. S. diagram in Fig.~\ref{feynman}. 
Let us rewrite $s_{21}$ in Eq.~(\ref{secondperturbation}) in terms of
the complete set of eigenstates $H_0$, including the complete set of bound states $\vert \phi_m \rangle$
and the continuum of the free-electron states. We get,
\begin{widetext}
\begin{align}
s_{21}&=\Big(\frac{-i}{\hbar}\Big)^2\sum_m\int dk_{m'}\int_{-\infty}^{+\infty} 
dt_1\int_{-\infty}^{t_1} dt_2
\Big\langle \phi_f \Big\vert {\cal V}_2^{\rm I}(t_1)\Big\vert \phi_m 
\Big\rangle \Big\langle
\phi_m\Big\vert {\cal V}^{\rm I}_1(t_2) \Big\vert\phi_i
\Big\rangle,
\label{ampiltude1}
\end{align}
\end{widetext}
where $\vert \phi_m(r_1,r_2) \rangle$ is $\vert \phi_m \rangle=\vert \psi_m(r_1) \rangle e^{i{\bm k}_{m'}\cdot{\bm r}_2}/(2\pi)^{3/2}$. Inserting Eq.~(\ref{interpoten}) into Eq.~(\ref{ampiltude1}) results in
\begin{align}
s_{21}&=\frac{1}{2\pi^6}\Big(\frac{-i}{\hbar}\Big)^2\sum_m\int 
dk_{m'}\int_{-\infty}^{+\infty}dt_1\int_{-\infty}^{t_1} dt_2\nonumber\\
&\Big \langle \psi_f(r_1)e^{i{\bm k}_f\cdot {\bm r}_2}
\Big\vert {\cal V}_2 \Big\vert \psi_m(r_1)e^{i{\bm k}_{m'}
\cdot {\bm r}_2}\Big\rangle\times e^{\frac{i}{\hbar}(E_f-E_{mm'})t_2}\times\nonumber\\
\times &\Big \langle \psi_m(r'_1)e^{i{\bm k}_{m'}
\cdot {\bm r'}_2}\Big\vert {\cal V}_1 
\Big\vert \psi_i(r'_1)e^{i{\bm k}_{i}\cdot {\bm r'}_1}
\Big\rangle\times e^{\frac{i}{\hbar}(E_{mm'}-E_{i})t_1}
\label{amplitude2}
\end{align}
Plugging Eq.~(\ref{potentiala}) in Eq.~(\ref{amplitude2}) and performing the integration with respect to $t_2$ 
yields  the following expression for $s_{21}$,
\begin{align}
s_{21}&=\frac{eE}{(2\pi^2)}\Big(\frac{-i}{\hbar} \Big)^2\sum_m
\int dk_{m'}\int_{-\infty}^{\infty} dt_1\times\nonumber\\
&\times\Big(\frac{i\hbar}{E_{mm'}-E_{f}} \Big)
\times e^{\eta t_1}e^{\frac{i}{\hbar}(E_f-E_i)t_1}\sin(\omega_{\rm L}t_1)\times\nonumber\\
& \times\Big \langle \psi_f(r_1)e^{i{\bm k}_f
\cdot {\bm r}_2}\Big\vert {\cal V}_2 
\Big\vert \psi_m(r_1)e^{i{\bm k}_{m'}\cdot {\bm r}_2}
\Big\rangle\times\nonumber\\
&\times\Big \langle \psi_m(r'_1)e^{i{\bm k}_{m'}
\cdot {\bm r'}_2}\Big\vert V_1 
\Big\vert \psi_i(r'_1)e^{i{\bm k}_{i}\cdot {\bm r'}_1}\Big\rangle.
\label{amplitude3}
\end{align}
We convert the $\sin$-term to the exponential terms $e^{-i\omega_{\rm L}t}$ and $e^{i\omega_{\rm L}t}$ in 
Eq.~(\ref{amplitude3}). The expression including $e^{-i\omega_{\rm L}t}$ represents the absorption 
process in the joint electron-photon collision of Eq.~(\ref{reaction}), while the other one 
including $e^{i\omega_{\rm L}t}$ denotes the emission process associated to Eq.~(\ref{reaction}). 
In our calculation, we keep only the absorption part in Eq.~(\ref{amplitude3}), {\it i.e.}, the 
term involving $e^{-i\omega_{\rm L}t}$. For emission part, it is sufficient to change the laser 
frequency sign from $-$ to $+$ in Eqs.~(\ref{amplitude3a})-(\ref{amplitude4}). After performing 
the time integration $t_1$, $s_{21}$ reads
\begin{subequations}
\begin{align}
s_{21}&=-2\pi i \delta(E_f-E_i-\hbar\omega_{\rm L}-i\hbar \eta) A_{21},
\label{amplitude3a}
\end{align}

\begin{align}
A_{21}&=\frac{-eE}{(2\pi)^6}\frac{1}{2i\hbar}\sum_{m}\int dk_{m'}
\Big \langle \psi_f(r_1) e^{i{\bm k}_f\cdot {\bm r}_2}\Big\vert {\cal V}_2 \Big\vert
\psi_m(r_1)e^{i{\bm k}_{m'}\cdot {\bm r}_2}\Big\rangle\nonumber\\
&\frac{1}{E_f-E_{mm'}}\Big \langle \psi_m(r'_1)e^{i{\bm k}_{m'}\cdot {\bm r'}_2}
\Big\vert V_1 \Big\vert \psi_i(r'_1)e^{i{\bm k}_{i}\cdot {\bm r'}_2}\Big\rangle.
\label{amplitude4}
\end{align}
\label{amplitude4a}
\end{subequations}
In the following, we focus on the evaluation of the matrix elements appearing in Eq.~(\ref{amplitude4}).
To do so, we insert Eq.~(\ref{potentialb}) into the first matrix element of Eq.~(\ref{amplitude4}), which
reads
\begin{align}
 &\Big \langle \psi_f(r_1)
e^{i{\bm k}_f\cdot {\bm r}_2}\Big\vert {\cal V}_2 \Big\vert \psi_m(r_1)e^{i{\bm k}_{m'}\cdot
{\bm r}_2}\Big\rangle=\frac{4\pi e^2}{\vert {\bm k}_{m'}-{\bm k}_f\vert^2}\times\nonumber\\
&\Big \langle\psi_f(r_1)\Big \vert e^{i({\bm k}_{m'}-{\bm k}_f)\cdot {\bm r}_2}-1
\Big\vert \psi_m (r_1) \Big\rangle.
\label{matrixelement2}
\end{align}
Inserting Eq.~(\ref{potentiala}) into the  second matrix element of Eq.~(\ref{amplitude4}) yields
\begin{align}
&\Big \langle \psi_m(r'_1)e^{i{\bm k}_{m'}\cdot {\bm r'}_2}\Big\vert V_1 \Big\vert
\psi_i(r'_1)e^{i{\bm k}_{i}\cdot {\bm r'}_2}\Big\rangle=\nonumber\\
&=(2\pi)^3 eE\Bigg[\Big \langle \psi_{m}(r'_1)\Big\vert {\bm \epsilon}\cdot {\bm r}'_1\Big \vert
\psi_i(r'_1)\Big\rangle \delta^{3}({\bm k}_i-{\bm k}_{m'})+\nonumber\\
&\Big\langle \psi_m \Big \vert \psi_i\Big\rangle{\bm \epsilon}\cdot 
\frac{\partial}{i\partial{\bm q}'}\delta^3({\bm q}')\Bigg],
\label{matrixelement1}
\end{align}
where ${\bm q}'= {\bm k}_i-{\bm k}_{m'}$ and $\delta$ refers to the delta function. After substituting 
Eqs.~(\ref{matrixelement2}) and (\ref{matrixelement1}) in Eq.~(\ref{amplitude4}), the amplitude $A_{21}$ 
reads 
 \begin{align}
 A_{21}&=\frac{-eE}{(2\pi)^3}\frac{1}{2i\hbar}\sum_m\int dk_{m'}\frac{4\pi e^2}{\vert
 {\bm k}_{m'}-{\bm k}_f\vert ^2}\frac{1}{E_f-E_{mm'}}\nonumber\\
 &\Bigg[\Big\langle \psi_f(r_1)\Big\vert
 e^{i({\bm k}_{m'}-{\bm k}_f)\cdot {\bm r}_1}-1\Big \vert \psi_m(r_1)\Big\rangle
 \Big\langle \psi_m(r'_1)\Big \vert {\bm \epsilon}\cdot {\bm r}'_1\Big \vert
 \psi_i(r'_1)\Big\rangle\times
\delta^{3}({\bm k}_i-{\bm k}_{m'})+\nonumber\\
&+\Big\langle \psi_f(r_1)\Big\vert e^{i({\bm k}_{m'}-{\bm k}_f)\cdot
 {\bm r}_1}-1\Big \vert \psi_m(r_1)\Big\rangle\Big\langle \psi_m(r_1)\Big \vert
 \psi_i(r_1)\Big\rangle{\bm \epsilon}\cdot 
\frac{\partial \delta^3({\bm q}')}{i\partial {\bm q}'}\Bigg].
 \label{amplitude6}
 \end{align}
Eq.~(\ref{amplitude6}) is simplified by using the following identity
\begin{align}
\int \frac{\partial \delta(x)}{\partial x}\phi(x) dx=
-\int \delta(x) \frac{\partial \phi (x)}{\partial x} dx.
\label{identity}
\end{align}
Thus, the amplitude $A_{21}$ is 
\begin{align}
A_{21}&= \frac{-eE}{(2\pi)^3}\frac{1}{2i\hbar}\frac{4\pi e^2}{q^2}
\sum_m \Big\langle \psi_f(r_1)\Big\vert e^{i{\bm q}\cdot
{\bm r}_1}-1\Big \vert \psi_m(r_1)\Big\rangle\times\nonumber\\
&\times\frac{1}{E_f-E_{mi}}\Big\langle \psi_m(r'_1)
\Big \vert {\bm \epsilon}\cdot {\bm r}'_1\Big \vert
\psi_i(r'_1)\Big\rangle,
\label{amplitude7}
\end{align}
where the momentum transfer ${\bm q}= {\bm k}_i-{\bm k}_f$. By using the conservation energy [see 
Eq.~(\ref{amplitude3a})], the energy denominator in Eq.~(\ref{amplitude7}) reads 
\begin{align}
 E_f-E_{mi}={\cal E}_i-{\cal E}_m+\hbar \omega_{\rm L} +i\hbar \eta.
\label{energy denominaror}
\end{align}
Thus, the transition amplitude $A_{21}$ is,
\begin{align}
A_{21}&= \frac{-eE}{(2\pi)^3}\frac{1}{2i\hbar}\frac{4\pi e^2}{q^2}
\sum_m \frac{\Big\langle \psi_f(r_1)\Big\vert e^{i{\bm q}\cdot
{\bm r}_1}-1\Big \vert \psi_m(r_1)\Big\rangle \Big\langle \psi_m(r'_1)
\Big \vert {\bm \epsilon}\cdot {\bm r}'_1\Big \vert
\psi_i(r'_1)\Big\rangle}{{\cal E}_i-{\cal E}_m+\hbar \omega_{\rm L}+i\hbar \eta}
\label{amplitude8}
\end{align}
A more simplified result can be obtained if we use the Bethe-Born approximation, in which only the 
leading term of the exponential expression $e^{i{\bm q}\cdot{\bm r}_1} $ is taken into 
account~($e^{i{\bm q}\cdot{\bm r}_1} \simeq 1+i{\bm q} \cdot{\bm r}_1$). Therefore, the transition
amplitude reads 
\begin{align}
 A_{21}&= \frac{eE}{(2\pi)^3}\frac{1}{2\hbar}\frac{4\pi e^2}{q^2}
\sum_m
\frac{\Big\langle \psi_f(r_1)\Big\vert {\bm q}\cdot {\bm r}_1\Big \vert
\psi_m(r_1)\Big\rangle \Big\langle \psi_m(r'_1)\Big \vert {\bm \epsilon}\cdot {\bm r}'_1\Big \vert
\psi_i(r'_1)\Big\rangle}{{\cal E}_i-{\cal E}_m+\hbar \omega_{\rm L}+i\hbar \eta}.
\label{amplitude9}
\end{align}
As mentioned, the same computational method can be utilized to lead the transition
amplitude $A_{12}$. Thus, $s_{12}$ corresponding to the L. H. S. diagram in 
Fig.~\ref{feynman} reads,
\begin{subequations}
\begin{align}
 s_{12}&=-2\pi i \delta(E_f-E_i-\hbar \omega_{\rm L}-i\hbar\eta)A_{12}
\label{amplitude10a}
\end{align}
\begin{align}
A_{12}&=\frac{eE}{(2\pi)^3}\frac{1}{2\hbar}\frac{4\pi e^2}{q^2}
\sum_m
\frac{\Big\langle \psi_f(r_1)\Big\vert {\bm \epsilon}\cdot {\bm r}_1\Big \vert
\psi_m(r_1)\Big\rangle \Big\langle \psi_m(r'_1)\Big \vert {\bm q}\cdot {\bm r}'_1 \Big \vert
\psi_i(r'_1)\Big\rangle}{{\cal E}_f-{\cal E}_m-\hbar \omega_{\rm L}-i\hbar \eta}
\label{amplitude10b}
\end{align}
\label{amplitude10}
\end{subequations}
Therefore, $s_{if}$ for the process introduced in Eq.~(\ref{reaction}) is
\begin{align}
s_{if}&=-2\pi i \delta(E_f-E_i-\hbar \omega_{\rm L}-i\hbar\eta)A_{if},
\label{amplitude11a}
\end{align}
where $A_{if}$ can be obtained by summing Eqs.~(\ref{amplitude8}) and (\ref{amplitude10b}), 
\begin{subequations}
\begin{align}
 A_{if}=A_{12}+A_{21}
\label{amplitude11c}
\end{align}

\begin{align}
A_{if}=\frac{e^2E}{(2\pi)^2 q^2\hbar}
\sum_m&\Bigg\{
\frac{\Big\langle \psi_f(r_1)\Big\vert {\bm \epsilon}\cdot {\bm r}_1\Big \vert
\psi_m(r_1)\Big\rangle \Big\langle \psi_m(r'_1)\Big \vert {\bm q}\cdot {\bm r}'_1 \Big \vert
\psi_i(r'_1)\Big\rangle}{{\cal E}_f-{\cal E}_m-\hbar \omega_{\rm L}-i\hbar \eta}+\nonumber\\
&+\frac{\Big\langle \psi_f(r_1)\Big\vert {\bm q}\cdot {\bm r}_1\Big \vert
\psi_m(r_1)\Big\rangle \Big\langle \psi_m(r'_1)\Big \vert {\bm \epsilon}\cdot {\bm r}'_1\Big \vert
\psi_i(r'_1)\Big\rangle}{{\cal E}_i-{\cal E}_m+\hbar \omega_{\rm L}+i\hbar \eta}\Bigg\}.
\label{amplitude11b}
\end{align}
\label{amplitude11}
\end{subequations}
In Eq.~(\ref{amplitude11b}), the adiabatic switching of the interaction allows to consider $\eta \rightarrow 0$. 
The scattering amplitude of Eq.~(\ref{amplitude11b}) in a concise form thus reads (in atomic units, 
$\hbar=m_e=e=1$)
\begin{align}
 A_{if}&=\frac{E}{(2\pi q)^2}\Bigg\{ \Big\langle 
\psi_f(r_1) \Big \vert {\bm \epsilon}\cdot {\bm r}_1
 \frac{1}{{\cal E}_f-H_0-\hbar \omega_{\rm L}}{\bm q}
\cdot {\bm r}'_1\Big \vert \psi_i(r'_1) \Big \rangle\nonumber\\
&+\Big\langle \psi_f(r_1) \Big \vert {\bm q}
\cdot {\bm r}_1 \frac{1}{{\cal E}_i-H_0+\hbar 
\omega_{\rm L}}{\bm \epsilon}\cdot {\bm r}'_1
\Big \vert \psi_i(r'_1) \Big \rangle\Bigg\},
\label{finalamplitude}
\end{align}
It is important to note that Eq.~(\ref{finalamplitude}) works for both $S-S$ and $S-D$ transitions, 
while the authors in Ref.~\cite{NrFf1978} obtained the same result for transitions between 
states with the same parity~($S-S$ transitions). 

In Sec.~\ref{sstransition}, we consider the scattering amplitude $A_{if}$ of Eq.~(\ref{finalamplitude}) 
only for the $S-S$ transitions, where the Sturmian representation of the Coulomb Green's function is employed 
to evaluate analytically the matrix elements appearing in Eq.~(\ref{finalamplitude}). 

\subsection{Scattering amplitude for the $S-S$ transition}
\label{sstransition}
In this section, we use the Sturmian representation of the non-relativistic Coulomb Green's function 
in Eq.~(\ref{finalamplitude}) to get an explicit form for the scattering amplitude of the $S-S$ transition. 
The Schr\"{o}dinger-Coulomb Green function is defined~\cite{RsGd1991},
\begin{align}
 \frac{1}{{\cal E}_{i/f}-H_0\pm\hbar \omega_{\rm L}}&=
\sum_{l,m}g_{l}(\nu_{i/f},r,r') Y_{lm}(\Omega)Y^{*}_{lm}(\Omega'),
\label{coulombgreen}
\end{align}
where the summation is over all possible intermediate states (discrete and continuum states). The symbol 
$ Y_{lm}$ is designated for the spherical harmonics defined in the spherical coordinate system. The function
$g_{l}(\nu,r,r')$ reads~\cite{RsGd1991}
\begin{align}
 g_{l}(\nu_{i/f}r,r')&=\frac{2m}{\hbar^2}\left(\frac{2}{a_{\rm B}\nu}\right)^{2l+1}(rr')^{l}
e^{-(r+r')/a_{\rm B}\nu}\nonumber\\
&\times\sum_{k=0}^{\infty}\frac{k!}{(2l+1+k)!(l+1+k-\nu)}L_k^{2l+1}\left(\frac{2r}{a_{\rm B}\nu}\right)
L_k^{2l+1}\left(\frac{2r'}{a_{\rm B}\nu}\right),
\label{coulombgreen1}
\end{align}
where $L_k^{2l+1}(x)$ refers to the Laguerre polynomial. In the above equation, $\hbar$ and $a_{\rm B}$ 
denote Planck constant and Bohr radius, respectively. Note that for simplicity, we drop indexes of $r_1$ 
and $r'_1$ in the rest of our calculation. We exploit the energy parameterization~\cite{RsGd1991}
\begin{align}
 \nu=\frac{Z\hbar}{a_{\rm B}}\sqrt{-\frac{1}{2m_e E}},
\label{parametrization}
\end{align}
where $E=-(\alpha Z)^2m_ec^2/2n^2$. In Eq.~(\ref{coulombgreen1}), one uses the identity 
\begin{align}
 L_k^{2l+1}(x)=\frac{\Gamma(2l+2+k)}{(2l+2)\Gamma(k+1)}\:_1F_1(-k,2l+2,x)
\label{indentity}
\end{align}
and rewrite Eq.(\ref{coulombgreen1}) in terms of the confluent Hypergeometric function $\:_1F_1$ 
and $\Gamma$ function. Thus, $g_{l}(\nu,r,r')$ reads,
\begin{align}
g_{l}(\nu,r,r')&=\frac{2m}{\hbar^2}\left(\frac{2}{a_{\rm B}\nu}\right)^{2l+1}(rr')^{l}
e^{-(r+r')/a_{\rm B}\nu}\nonumber\\
&\times\sum_{k=0}^{\infty} \frac{\Gamma(2l+2+k)[\Gamma(2l+2)]^{-2}}{(l+1+k-\nu)\Gamma(k+1)}
\:_1F_1(-k,2l+2,\frac{2r}{a_{\rm B}\nu})\:_1F_1(-k,2l+2,\frac{2r'}{a_{\rm B}\nu})
\label{coulombgreen2}
\end{align}  
In this position, one can plug Eq.~(\ref{coulombgreen}) in Eq.~(\ref{finalamplitude}) and 
performed the required angular integrations. The scattering amplitude of the  $S-S$ 
transitions induced by SEPE process thus is 
\begin{subequations}
\begin{align}
A_{if}&=\frac{E}{(2\pi)q^2}({\bm q}\cdot {\bm \epsilon})\frac{a_{if}}{3}
\label{amplitudeA}
\end{align}

\begin{align}
a_{if}&=\int_0^\infty r^3r'^3 R_{n_f,0}(r) R_{n_i,0}(r')[g_{1}(\nu_f,r,r')+g_{1}(\nu_i,r,r')]dr dr' \nonumber\\
&=\frac{4}{9\nu^3}\sum_{k=0}^\infty \frac{\Gamma(k+4)}{\Gamma(k+1)(2+k-\nu)}
[F_{n_f}(\nu_f)F_{n_i}(\nu_f)+ F_{n_f}(\nu_i)F_{n_i}(\nu_i)]
\label{amplitudeB}
\end{align}

\begin{align}
F_{n_f/n_i}(\nu_{f/i})&=\int_{k=0}^\infty r^4 e^{-\frac{r}{a_{\rm B}\nu}}\:_1F_1 (-k,4,\frac{2r}{a_{\rm B}, \nu_{f/i}})R_{n_f/n_i,0}(r) dr
\label{amplitudeC}
\end{align}
\label{amplitudes}
\end{subequations}
where $R_{n_f,0}(r)$ and $R_{n_i,0}(r)$ are the radial part of the wave function in the ground and excited 
states. In this investigation, the angular momentum quantum number $l$ of the virtual states in the 
propagator for the $S-S$ transition is $P$ states. 

In Sec.~\ref{total_cross_section}, we shall employ the transition amplitude of Eq.~(\ref{amplitudes}) 
in order to evaluate the excitation cross section of the SEPE process of Eq.~(\ref{reaction}).

\begin{figure}[htb]
\begin{center}
\begin{minipage}{0.7\linewidth}
\begin{center}
\includegraphics[width=1.1\linewidth]{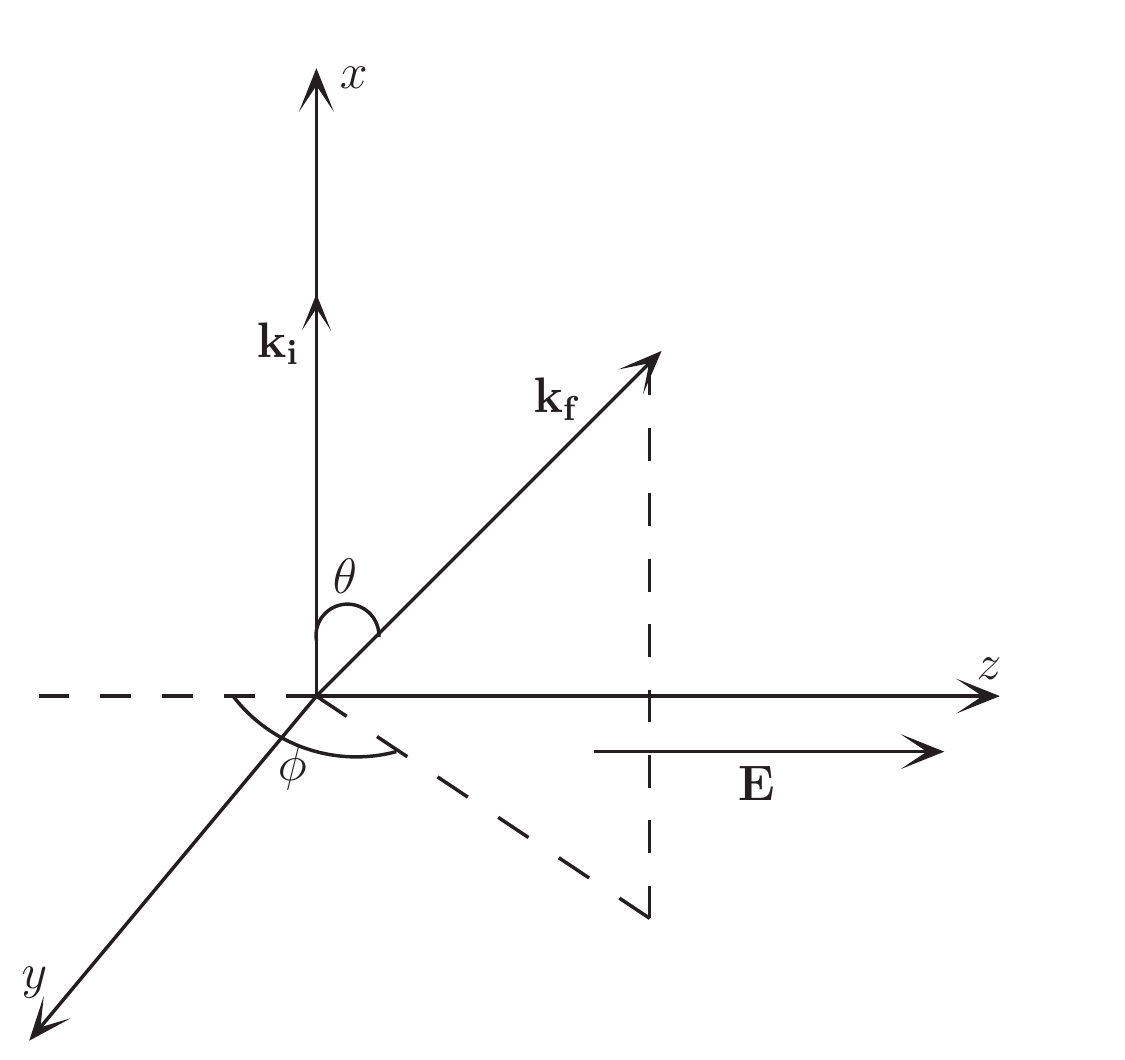}
\caption{\label{coor}
The electric field ${\bm E}$ in the direction of the quantization
axis $z$ in the unrotated coordinate system.}
\end{center}
\end{minipage}
\end{center}
\end{figure}

\subsection{ Total cross section}
\label{total_cross_section}
This section is dedicated to the evaluation of the excitation cross section of the $S-S$ transitions  
induced by the SEPE process of Eq.(\ref{reaction}). The partial cross section in terms of the transition 
amplitude is defined as follows~\cite{Cj1975,Jj1994}
\begin{align}
 \frac{d\sigma}{d\Omega}=(2\pi)^4\frac{k_f}{k_i}\Big\vert A_{if}\Big\vert^2.
\label{diffcrosssection}
\end{align} 
Substituting Eq.~(\ref{amplitudes}) into Eq.~(\ref{diffcrosssection}) yields 
\begin{align}
\frac{1}{I_{\rm L}}\frac{d\sigma}{d\Omega}=8\pi \alpha \frac{k_f}{k_i}
\frac{(\bm{q}\cdot\bm{\epsilon})^2}{q^4}
\left \vert \frac{a_{if}}{3}\right\vert^2,
\label{diffcrosssection1}
\end{align}
where $I_{\rm L}=E^2/8\pi \alpha$ and $\alpha$ is fine structure constant. In the present work, it is 
assumed that the electric field is the quantization axis~[see Fig.~\ref{coor}]. Since it is more 
convenient to obtain the total cross section in the coordinate system that the transfer momentum 
${\bm q}$ is the quantization axis, it is required to rotate the coordinate system such a way that ${\bm q}$ 
is the quantization axis in its final status. In this new coordinate system, the components of the 
electric field unit vector $(\epsilon_0, \epsilon_{\pm 1})$ in the spherical coordinate system are 
\begin{align}
 \epsilon_0&=\frac{k_f\cos(\theta_{\rm p}-\theta)-k_i\cos \theta_{\rm p}}{q}\nonumber\\
 \epsilon_{\pm 1}&=\frac{-i}{\sqrt{2}}
\frac{\sin\theta_{\rm p}(k_f\cos\theta-k_i)-
k_f\cos\theta_{\rm p}\sin\theta}{q},
\label{polaricompon}
\end{align}
where $\theta$ is the scattering angle. $\theta_{\rm p}$ refers to the angle between the electric
field unit vector (polarization vector $\bm{\epsilon}$) and the initial momentum of the incoming electron. 

Based on the orientation of the electric filed with unit vector ${\bm \epsilon}$ with respect to the 
initial momentum of the electron ${\bm k}_i$, one can obtain the total cross section for the two 
special geometries of the joint electron-photon scattering introduced in Eq.~(\ref{reaction}). 
Here, we present the analytical results of the total excitation cross sections for the following geometries:
\begin{itemize}
 \item $\bm{\epsilon}\bot \bm{k}_i$:   We consider the case that the electric field is perpendicular 
to the initial momentum $k_i$~($\theta_{\rm p}=\pi/2$). The total cross section for this geometry can 
be calculated by performing the angular integration in Eq.~(\ref{diffcrosssection1}) and reads
\begin{align}
 \frac{\sigma_\bot}{I_{\rm L}}=8\pi \alpha \frac{k_f}{k_i}A_{\bot}
\left \vert \frac{a_{if}}{3}\right\vert^2,
\label{totalperpen}
\end{align}
 where $A_{\bot}$ is
\begin{align}
 A_{\perp}=\frac{\pi}{k_i^2}\Big[\frac{1}{\chi}\ln\Big(\frac{1+\chi}{1-\chi}\Big)-2\Big],
\label{anugualrperpen}
\end{align}
and $\chi$ is defined
\begin{align}
 \chi=\frac{2k_ik_f}{k_i^2+k_f^2}
\label{kapa}
\end{align}
Here, we should add that there is a deviation between the expression of $A_{\perp}$ and the 
one in Ref.~\cite{NrFf1978},
\begin{align}
 A'_{\perp}=\frac{\pi}{2k_i^2}\Big[\frac{1}{\chi}\ln\Big(\frac{1+\chi}{1-\chi}\Big)\Big].
\label{anugualrperpenref}
\end{align}
We have checked our result in Eq.~(\ref{anugualrperpen}). It seems that there is a mistake
in Ref.~\cite{NrFf1978}, see also Appendix~\ref{appendixB}.
\item $\bm{\epsilon}\parallel \bm{k}_i$:  We envisage the case that the polarization vector 
$\bm{\epsilon}$ is parallel to the initial momentum $k_i$ of the electron ($\theta_{\rm p}=0$). 
This assumption leads the following result for the total cross section,
\begin{align}
 \frac{\sigma_\parallel}{I_{\rm L}}=8\pi \alpha \frac{k_f}{k_i}A_{\parallel}
\left \vert \frac{a_{if}}{3}\right\vert^2,
\label{totalpararel}
\end{align}
where $A_\parallel$ reads
\begin{align}
 A_\parallel=\frac{\pi}{k_i^2}\Big[\Big(\frac{1}{\chi}-
\frac{1}{\varrho}\Big)\ln\Big(\frac{1+\chi}{1-\chi}\Big)+2\Big],
\label{angularpara}
\end{align}
and $\varrho= k_i/k_f$.
\end{itemize}
\begin{table}[t]
\caption{\label{tab1} Total cross sections in the units of $\pi a_0^2$ for the SEPE process of Eq.~(\ref{reaction}). 
These values are calculated based on Eq.~(\ref{totalperpen}) for $\bm{\epsilon}\bot \bm{k}_i$ 
geometry. Here, the laser intensity and frequency are $I_{\rm L}=10^{13}~{\rm W/cm^2}$ 
and $\omega_{\rm L}= 1.17~{\rm eV}$, respectively. We also present total excitation cross 
sections in absence of laser field. } 
\begin{tabular}{lll|ll}
\hline\hline
 &\multicolumn{2}{l}{~~~~$E=100~ {\rm eV}$}&\multicolumn{2}{l}{~~~~$E=200~ {\rm eV}$}\\
\cline{2-5}
Transitions &~~Laser on&Laser off&~~Laser on&Laser off\\
\hline\hline
$1S\rightarrow 2S$&$2.6610\times 10^{-1}$&$5.77\times 10^{-2}\footnotemark[1]$&$1.7173\times 10^{-1}$&$2.95\times 10^{-2}\footnotemark[1]$\\
$1S\rightarrow 3S$&$2.8582\times 10^{-1}$&$1.20\times 10^{-2}\footnotemark[2]$&$1.8869\times 10^{-1}$&$5.99\times 10^{-3}\footnotemark[2]$\\
$1S\rightarrow 4S$&$1.9891\times 10^{-2}$&~~~~--&$1.0323\times 10^{-2}$&~~~~--\\
$1S\rightarrow 5S$&$1.2356\times 10^{-3}$&~~~~--&$8.2514\times 10^{-4}$&~~~~--\\
\hline\hline
\end{tabular}
\footnotetext[1]{These values are taken from Ref.~\cite{AkWcPg1976}} 
\footnotetext[2]{These values are taken from Ref.~\cite{AkJe1966}}
\end{table}

\section{Numerical results and discussion}
\label{neumeri}
This section is dedicated to the numerical evaluations of the cross section of the SEPE process 
of Eq.~(\ref{reaction}) for $1S-nS$ transitions, $n\in \lbrace 2,3,4,5\rbrace$. Our method is
relied on the Schr\"{o}dinger Coulomb Green function (Sturmian representation of the Green function),
which was described in Sec.~(\ref{sstransition}). For this investigation, we take into account  
two different geometries discussed in the previous section. In the numerical calculations presented 
in this work, we consider the Nd:YAG laser. For the laser frequency and intensity, we use the 
numerical values introduced in Ref.~\cite{NrFf1976}. According to this reference, the laser 
frequency $\omega_{\rm L}$ and the laser intensity $I_{\rm L}$ are $1.17~{\rm eV}$ and 
$10^{13}~{\rm W/cm^2}$, respectively.

\subsection{Cross section for $\epsilon \perp k_i $ geometry}
\label{perpendicular_geometry}
The total cross section for the geometry $\bm{\epsilon} \perp \bm{k}_i $ is obtained by using 
Eq.~(\ref{totalperpen}). In this calculation, it is required that the radial integrals and the 
summation appearing in Eq.~(\ref{amplitudes}) are calculated. We perform analytically the radial 
integrals and the summations for $1S-nS$ transitions( $,n\in \{2,3,4,5\}$) induced by the SEPE process 
in the Hydrogen atom in appendix~\ref{appendixA}. We emphasize that both discrete and continuum 
states in the summation appearing in Eq.~(\ref{amplitudes}) are treated in our calculations. In 
table~\ref{tab1}, for these transitions, the total cross sections are obtained by treating a fast 
electron with initial energies $E_i=100$ and $200$~${\rm eV}$. We compare one of our results with the 
available one in the literature~\cite{NrFf1976}. We have found some inconsistencies between our result 
and the one in the literature. We have devoted appendix~\ref{appendixB} in order to deal with these 
inconsistencies. In Table~\ref{tab1}, we also compare some of our results with the excitation cross 
sections of the inelastic electron-Hydrogen atom scattering in the free field, which were appeared in 
literature, see {\it e.g.} Refs.~\cite{AkWcPg1976,AkJe1966}. As we expect, one or two order(s) of 
magnitude the excitation cross sections induced by the SEPE process is nearly larger than those in the 
free field, see also Table~\ref{tab1}.

In the following of our investigation, the behavior of the differential cross section as a function of the 
scattering angle $\theta$ is inspected . It is obtained by 
taking an integral over the azimuthal angle $\phi$ in Eq.~(\ref{diffcrosssection1}) leading to 
\begin{align}
 \frac{d\sigma}{d\theta}&=\int \frac{d\sigma}{d\Omega} d\phi=16\pi^2\alpha I_{\rm L}\frac{k_f}{k_i}
\left\vert \frac{a_{if}}{3}\right\vert^2
\left[\frac{k_f\sin\theta}{k_i^2(1+\frac{k_f^2}{k_i^2}-2\frac{k_f}{k_i}\cos\theta)}\right]^2.
\label{diffbha}
\end{align}

\begin{figure}[htb]
\begin{center}
\begin{minipage}{0.7\linewidth}
\begin{center}
\includegraphics[width=1.1\linewidth]{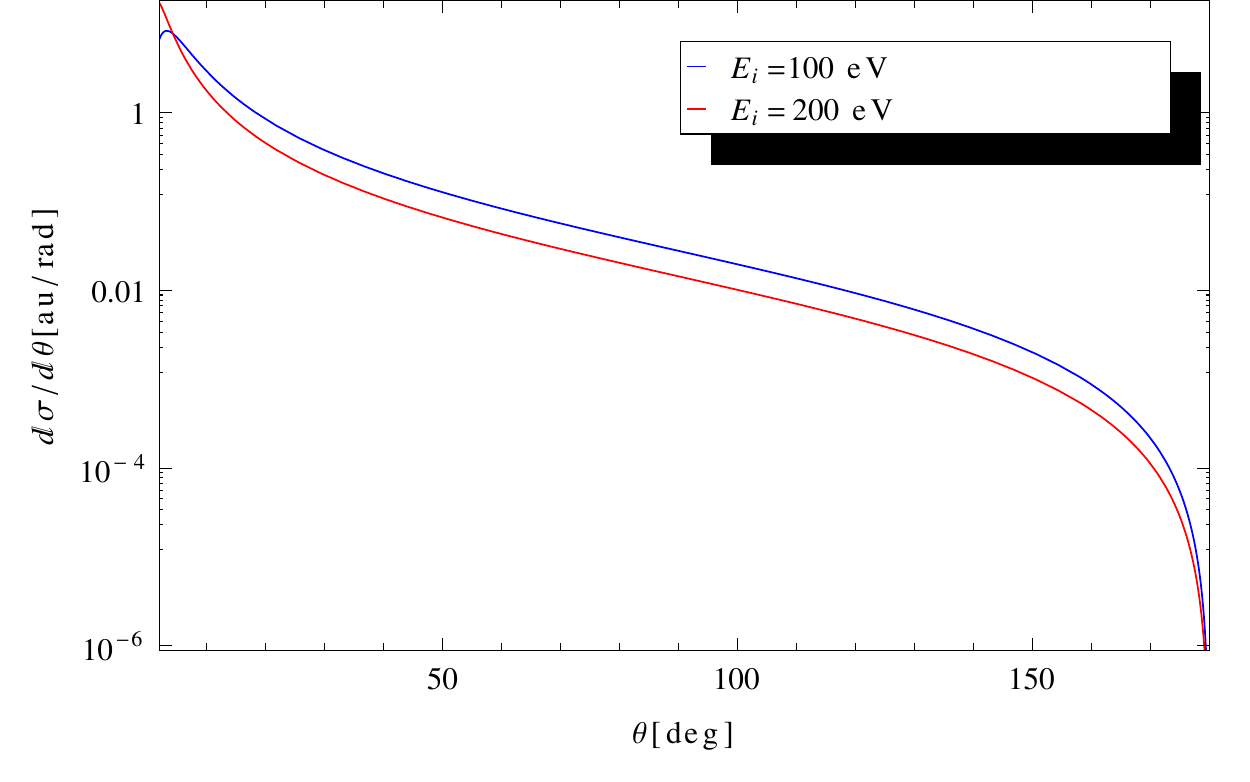}
\caption{The behavior of the differential cross section as a function of 
the scattering angle $\theta$ for the $1S-3S$ transition based on Eq.(\ref{diffbha}). 
The initial energy of the electron is assumed to be $100$ and $200$ ${\rm eV}$. 
The polarization vector of the laser field is perpendicular to the initial 
momentum of the electron.\label{fig5}} 
\end{center}
\end{minipage}
\end{center}
\end{figure}
In Fig.~\ref{fig5}, the behavior of the differential cross section as a function of the scattering angle
$\theta$ for the free electron with initial energies $E_i=100$ and $200$~${\rm eV}$ is compared for 
the particular case $1S-3S$ transition. We have also confirmed that the same behaviors 
for the other transitions $1S-nS$ with $n\in \{2,4,5\}$ as presented in Fig.~\ref{fig5}
for the $1S-3S$ transition are observed.

In order to interpret the behavior of the differential cross section of Fig.~(\ref{fig5}), we consider 
Eq.~(\ref{diffbha}) in the various ranges of the scattering angle $\theta$. First, we consider the scattering 
angle $\theta$ in the small and intermediate ranges. In these ranges, Eq.~(\ref{diffbha}) can be approximated
as follows,
 \begin{subequations}
\begin{align}
\frac{d\sigma}{d\theta}&=16\pi^2\alpha I_{\rm L}\frac{k_f}{k_i}
\left\vert\frac{a_{if}}{3}\right\vert^2 
 \frac{1}{\beta}\left(\frac{k_i}{k_f}\right)^3\theta^2, \qquad \beta\gg \frac{k_f}{k_i}\theta^2 
\label{diffasy1}
\end{align}
\begin{align}
\frac{d\sigma}{d\theta}&=16\pi^2\alpha I_{\rm L}\frac{k_f}{k_i}
\left\vert\frac{a_{if}}{3}\right\vert^2 
 \frac{1}{\beta}\left(\frac{k_f}{k_i^3}\right)\theta^{-2}, \qquad \beta\ll \frac{k_f}{k_i}\theta^2, 
\label{diffasy2}
\end{align}
\label{diffasy}
\end{subequations}
where $\beta=(1-k_f/k_i)^2$. Eq.~(\ref{diffasy1}) indicates that the maximum of $d\sigma/d\theta$ is 
taking place at very small scattering angle $\theta$ and an increase of the initial energy of the 
electron causes $(d\sigma/d\theta)_{max}$ to move toward zero. This is confirmed by the behavior of 
the maximum in Fig.~\ref{fig5}. For an intermediate scattering angle in Eq.~(\ref{diffasy2}), due 
to proportionality the differential cross section to $\theta^{-2}$, it is rapidly decreased with 
increasing the scattering angle $\theta$. This is corroborated by the behavior of $d\sigma/d\theta$ 
in the intermediate scattering angle of Fig.~\ref{fig5}. For large cross section, the differential 
cross section of Eq.~(\ref{diffbha}) is proportional to $1/\beta$. This means that $d\sigma/d\theta$ 
does not depend on $\theta$ at large scattering angle and it lowers with increasing the initial energy 
of the free electron $E_i$. This behavior of the differential cross section is substantiated by 
Fig.~\ref{fig5} at large scattering angle.
\begin{table}[h]
\caption{\label{tab2} Total cross sections in the units of $\pi a_0^2$ for SEPE process of Eq.~(\ref{reaction}). 
These values are calculated based on Eq.~(\ref{totalpararel}) for  $\epsilon\parallel k_i$
geometry. Here, the laser intensity and frequency are $I_{\rm L}=10^{13}~{\rm W/cm^2}$ 
and $\omega_{\rm L}= 1.17~{\rm eV}$, respectively.}
\begin{tabular}{lllll}
\hline\hline
Transitions &~~$E=100~ {\rm eV}$ &&~~$E=200~ {\rm eV}$&\\
\hline\hline
$1S\rightarrow 2S$&$1.1400\times 10^{-1}$& &$5.4721\times 10^{-2}$&\\
$1S\rightarrow 3S$&$1.3509\times 10^{-1}$& &$6.4666\times 10^{-2}$&\\
$1S\rightarrow 4S$&$9.7101\times 10^{-3}$& &$4.6444\times 10^{-3}$&\\
$1S\rightarrow 5S$&$6.1208\times 10^{-4}$& &$2.9267\times 10^{-4}$&\\
\hline\hline
\end{tabular}
 \end{table}
\subsection{Cross section for $\epsilon\parallel k_i$ geometry}
\label{parallel}
For this geometry, the total cross section of the SEPE process of Eq.~(\ref{reaction}) based on Eq.~(\ref{totalpararel}) for $1S-nS$ transitions ($n\in \lbrace 2,3,4,5\rbrace$) is calculated. We list the total 
cross sections in the Table~\ref{tab2}.  

In this geometry, the differential cross section as a function of the scattering angle $\theta$ is also 
investigated. It is 
\begin{align}
\frac{d\sigma}{d\theta}&=16\pi^2\alpha I_{\rm L}\frac{k_f}{k_i}
\left\vert\frac{a_{if}}{3}\right\vert^2
\left[\frac{1-(k_f/k_i)\cos\theta}{k_i^2(1+(k_f/k_i)^2-2(k_f/k_i)\cos\theta)}\right]^2.
\label{diffthetpara}
\end{align}

\begin{figure}[htb]
\begin{center}
\begin{minipage}{0.7\linewidth}
\begin{center}
\includegraphics[width=1.1\linewidth]{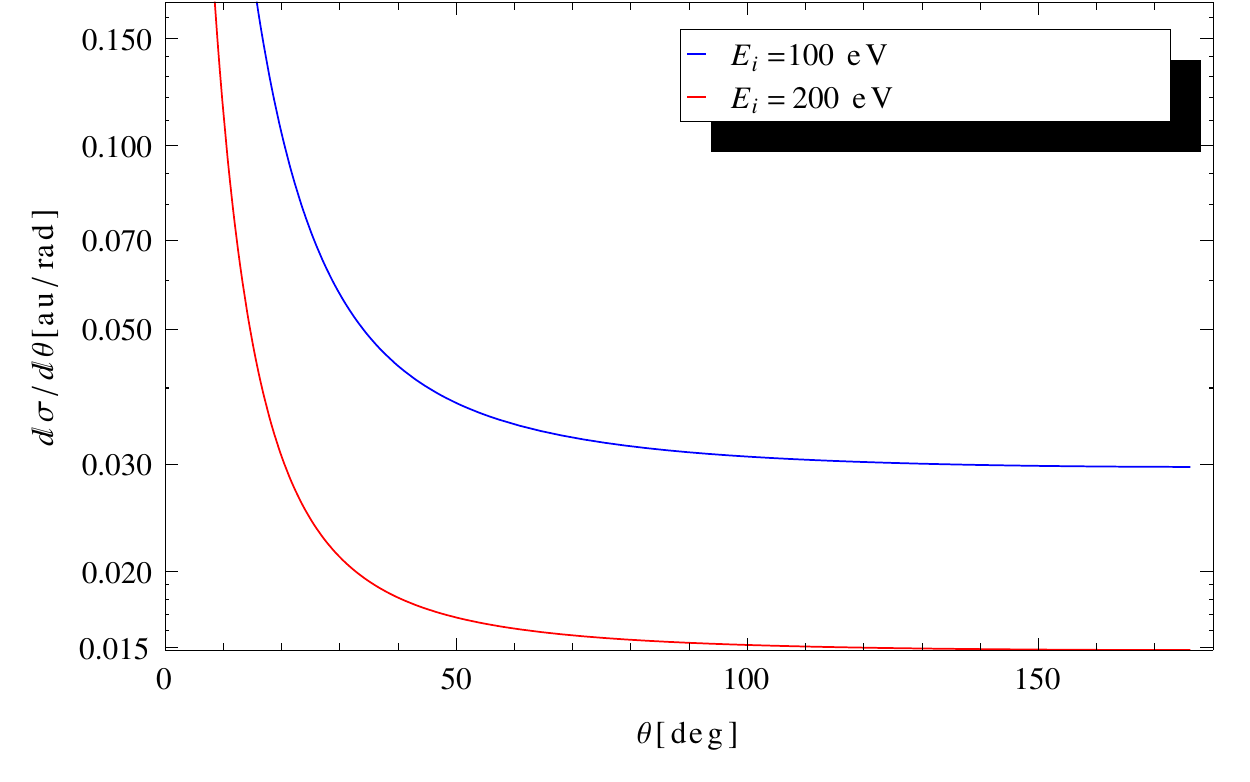}
\caption{The behavior of the differential cross section as a function of 
the scattering angle $\theta$ for the $1S-3S$ transition based on 
Eq.(\ref{diffthetpara}). The polarization vector of the laser field is 
parallel to the initial momentum of the electron\label{fig6}
} 
\end{center}
\end{minipage}
\end{center}
\end{figure}

We draw the differential cross section of Eq.~(\ref{diffthetpara}) as a function of $\theta$ for electrons 
with initial energies $E_i=100$ and $200~{\rm eV}$ in Fig.~\ref{fig6}. 
In this case, we also pay attention to the behavior of $d\sigma/d\theta$ in various regions of the scattering 
angle. At small and intermediate scattering angles $\theta$, the differential cross sections are
\begin{subequations}
\begin{align}
\frac{d\sigma}{d\theta}&=16\pi^2\alpha I_{\rm L}\frac{k_f}{k_i}
\left\vert\frac{a_{if}}{3}\right\vert^2 
 \frac{\beta^{-2}}{k_i^2}, \qquad \theta^2\ll \beta^2 < \beta 
\label{diffasy3}
\end{align}
\begin{align}
\frac{d\sigma}{d\theta}&=16\pi^2\alpha I_{\rm L}\frac{k_f}{k_i}
\left\vert\frac{a_{if}}{3}\right\vert^2 
 \frac{\beta^2\theta^{-4}}{k_i^2}, \qquad \beta^2\ll \theta^2<\beta, 
\label{diffasy4}
\end{align}
\label{diffasypar}
\end{subequations}
Eq.~(\ref{diffasy3}) indicates that the differential cross section is independent of the scattering angle 
$\theta$ at small scattering angle. An increasing energy of the incoming electron leads to a decreasing of the 
maximum of $d\sigma/d\theta$ at very small scattering angle $\theta$, which is also observed in Fig.~\ref{fig6}. 
For an intermediate scattering angle $\theta$, (see Eq.~(\ref{diffasy4})), the differential cross section is proportional to 
$\beta^2\theta^{-4}/k_i^2$, which describes a rapid decrease of $d\sigma/d\theta$ when the scattering angle is 
increased. For large scattering angle, it is proportional to  $1/2k_i^2$, which shows that the differential cross section 
is independent of the scattering angle and decreases when the energy of the incoming electron is increased.    
\begin{figure}[t]
\begin{center}
\begin{minipage}{0.7\linewidth}
\begin{center}
\includegraphics[width=1.1\linewidth]{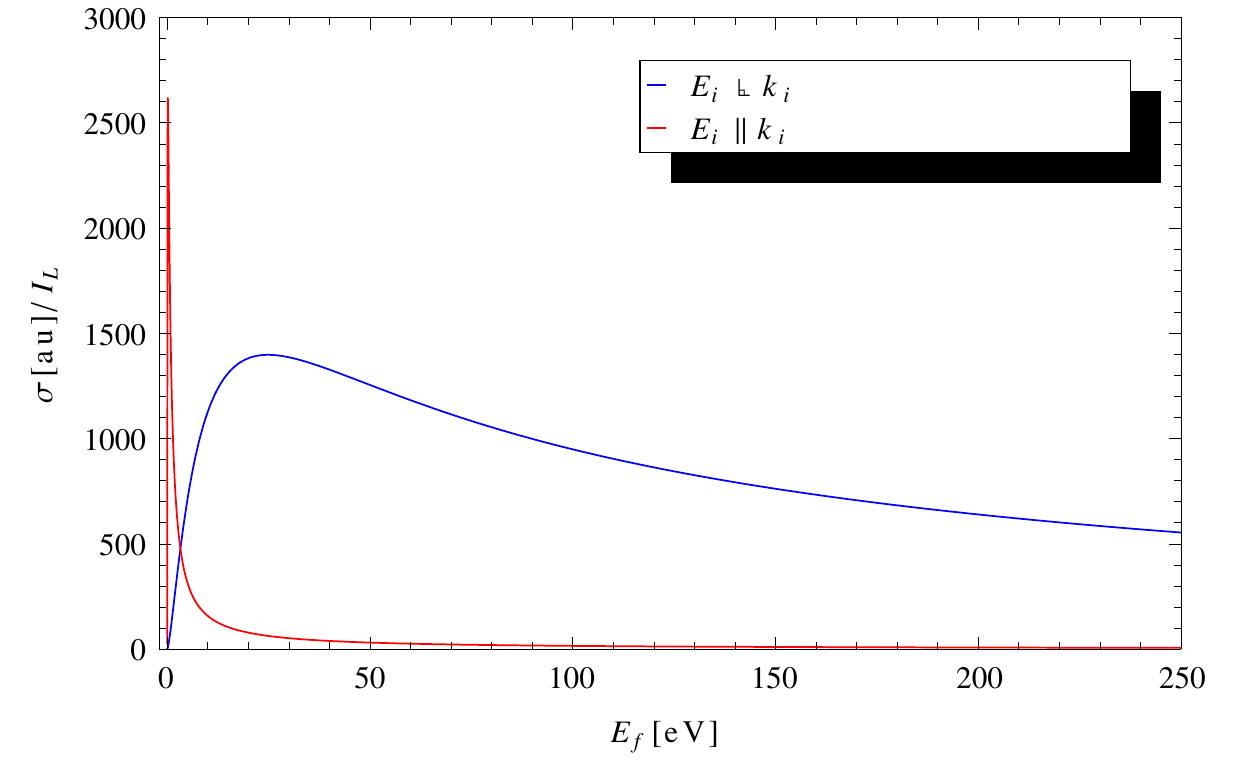}
\caption{The total cross section of the $1S-3S$ transition induced by the 
SEPE process as function of the final energy of the incident electron for 
two different geometries.\label{fig7}
} 
\end{center}
\end{minipage}
\end{center}
\end{figure}
\subsection{Effect of laser geometry on the cross section}
In the following of our investigations, we want to deal with effect of laser beam geometry on the 
total cross section in the SEPE process. In Fig.~\ref{fig7}, we depict the total cross section of 
the $1S-3S$ transition as a function of the final energy of the projectile for two different geometries
discussed in the previous sections. We have checked that there are the same behaviors for the excitation 
cross sections of $1S-nS$, with $n\in \{2,4,5\}$ as presented in Fig.~\ref{fig7} for the 
$1S-3S$ transition. 

In Fig.~\ref{fig7}, one can observe that the total cross section at low energy projectile tends to 
zero. Then, for $\epsilon\parallel k_i$ geometry, there is an extremely rapid increase in the total cross 
section and it starts to fall off very fast as $\ln E/E$. For $\epsilon \perp k_i $ geometry, a 
moderate increase in the total excitation cross section up to $\approx 25~{\it eV}$ is observed. Then, 
it starts to fall off as $\ln E/E$ at intermediate and high energies. The existence of logarithmic behavior 
in both cases is originated from the fact that the angular momentum is conserved in the atomic collisions 
in the presence of the laser field~\cite{NrFf1976}. At high energy projectile, where the Bethe-Born 
approximation works very well, the total cross section for the parallel geometry decreases more rapidly 
than the perpendicular geometry. As a final remark in Fig.~\ref{fig7} is that an intersection for 
excitation cross sections for these two different geometries is taken place at low energy part. This 
may be explained by the fact that the excitation cross section of the SEPE process depends on 
the projection of the momentum transfer $\bm{q}$ on the unit vector of the laser field $\bm{\epsilon}$, $(\bm {q} \cdot \bm{\epsilon})$. It is is always small at high and low energies for the parallel and perpendicular polarization, 
respectively, see Eqs.~(\ref{diffcrosssection1}) and (\ref{polaricompon}). As a consequence, the ratio of 
$\sigma_\parallel/\sigma_\bot$ is larger than unit for the low energy part of Fig.~\ref{fig7}, while 
$\sigma_\parallel/\sigma_\bot$ is less than unit at the high energy part.    
\section{Summary and outlook}
\label{summ}
The present study concerns with the theoretical analysis of the second-order SEPE process in the 
Hydrogen atom, where the photon energy of the laser field and the energy of the electron were assumed to 
be equal to the atomic state energy difference. The investigation was performed by 
considering the second-order contribution of Dyson's perturbation series for the evaluation 
of the excitation cross section induced by the SEPE process. Such a investigation for the 
second-order SEPE process was discussed earlier in the literature in relation to the appearance 
of resonance structures in the excitation cross section, where the transition amplitude relation was 
obtained only for the $S-S$ transition. In the current investigation, we used 
the dipole and Bethe-Born approximations, where the last one is an appropriate method when high-energy
projectile is considered in the SEPE process. The techniques were made possible to evaluate  
the transition amplitudes for the $S-S$, $S-D$ transitions, see Eq.~(\ref{finalamplitude}). Total 
cross sections in the closed forms for two different geometries described are derived in Sec.\ref{genral}. 
This leads to the characteristic dependency of the excitation cross section on the energy of the projectile 
(electron) and the scattering angle $\theta$. The total cross sections of $1S-nS, n\in \{2,3,4,5\}$ 
transitions are obtained. By comparing our results with the ones in the free field, we come to the conclusion 
that the Bethe-Born approximation generates acceptable results for the SEPE process in the range of 
high energy projectile (electron). 

It is natural to exploit the same idea for the numerical evaluation of the excitation cross section 
for the $S-D$ transitions induced by the SEPE process in the Hydrogen atom. 
This extension is left to the future work, and hoped to yield an improved investigation and better understanding of the 
SEPE process based on the Bethe-Born approximation.

\appendix
\section{Evaluation of the radial integrals}
\label{appendixA}
In this appendix, we give analytical results for radial integrals appearing in 
Eq.~(\ref{amplitudes}), which is a required step for the evaluation of the excitation cross sections of 
transitions considered in Secs.~\ref{perpendicular_geometry} and \ref{parallel}. In general, this 
radial integral reads
\begin{align}
a_{if}(\nu)&=\int_0^\infty r^3r'^3 R_{n_f,0}(r) R_{n_i,0}(r')g_{1}(\nu,r,r')dr dr',
\label{radialintegral_app}
\end{align}
where $g_{1}(\nu,r,r')$ is defined in Eq.~(\ref{coulombgreen2}). For each transition discussed in text, 
we obtain the following relations:
\begin{subequations}
\begin{align}
a_{1S-2S}(v)&=\frac{2^9\sqrt{2}}{3^5(v^2-4)^3(v^2-1)^2(v^2+3v+2)}\Upsilon_{1S-2S}(v),
\label{2Ssuma}
\end{align}
\begin{align}
 \Upsilon_{1S-2S}(v)&=256+384v-416v^2-816v^3-32v^4+360v^5-670v^6-5~073v^7-6~037v^8\nonumber\\
&-1~659v^9+65v^{10}+11~664v^8\:_2F_1\left(1,-v;1-v;\frac{(v-1)(v-2)}{(v+1)(v+2)}\right).
\label{2Ssumb}
\end{align}
\label{1Ssum}
\end{subequations}

\begin{subequations}
\begin{align}
 a_{1S-3S}(v)&=-\frac{3^3\sqrt{3}v^2}{2^6(v^2-9)^4(v^2-1)^2(v^2+4v+3)}\Upsilon_{1S-3S}(v),
\label{3Ssuma}
\end{align}

\begin{align}
 \Upsilon_{1S-3S}(v)&=-59~049-78~732v+91~854v^2+148~716v^3+18~225v^4-25~272v^5+303~588v^6\nonumber\\
&+1~740~312v^7+1~767~177v^8-42~492v^9-486~034v^{10}-104~132v^{11}+2~639v^{12}\nonumber\\
&+131~072v^{8}(-27+7v^2)\:_2F_1\left(1,-v;1-v;\frac{(v-1)(v-3)}{(v+1)(v+3)}\right).
\label{3Ssumb}
\end{align}
\label{3Ssum}
\end{subequations}

\begin{subequations}
\begin{align}
a_{1S-4S}(x)&=-\frac{2^{14}v^2}{3 \times 5^7 (v^2-16)^5(v^2-1)^2(v^2+5v+4)}\Upsilon_{1S-4S}(v),
\label{4Ssuma}
\end{align}
\begin{align}
\Upsilon_{1S-4S}(v)&=75~497~472 + 94~371~840v-115~605~504v^2-168~099~840v^3 -29~196~288v^4\nonumber\\
& +5~529~600v^5-492~011~520v^6-2~536~396~800v^7- 2~331~412~480v^8+479~833~600v^9\nonumber\\
&+925~767~588 v^{10}+94~751~085v^{11}-76~666~371v^{12}-13426~985 v^{13}+189~603v^{14}\nonumber\\
&+6~250~000v^8(768-288v^2+23v^4)\:_2F_1\left(1,-v;1-v;\frac{(v-1)(v-4)}{(v+1)(v+4)}\right).
\label{4Ssumb}
\end{align}
\end{subequations}

\begin{subequations}
\begin{align}
a_{1S-5S}(x)&=-\frac{2^25^4 \sqrt{5}v^2}{3^8 (v^2-25)^6(v^2-1)^2(v^2+6v+5)}\Upsilon_{1S-5S}(v),
\label{5Ssuma}
\end{align}

\begin{align}
\Upsilon_{1S-5S}(v)&=-1~220~703~125-1~464~843~750v+1~855~468~750v^2+2~519~531~250v^3\nonumber\\ 
 &+519~531~250v^4+119~531~250v^5+8~980~468~750v^6+43~557~656~250v^7\nonumber\\
 &+37~415~800~000v^8-12~363~971~250v^9-17~541~347~750v^{10}-471~962~250v^{11}\nonumber\\
&+2~213~557~150v^{12}+257~271~078v^{13}-81~752~630v^{14}-12~080~802v^{15}+109~381v^{16}\nonumber\\
&+1~679~616v^8(-46~875+20~625v^2-2~545v^4+91v^6)\:_2F_1\left(1,-v;1-v;\frac{(v-5)(v-1)}{(v+5)(v+1)}\right).
\label{5Ssumb}
\end{align}
\end{subequations}
Note that we use contiguous relations in derivations of above equations~\cite{Hb1953}.
\section{Comparison with literature source }
\label{appendixB}
This appendix is devoted to the discussion of the inconsistencies existing between our result for the excitation
cross section of the $1S-2S$ transition induced by the SEPE process and the literature source, see {\it e.g.}, 
Ref.~\cite{NrFf1976}. Since we want to find the main reason of inconsistencies between our result and 
the one presented in Ref.~\cite{NrFf1976}, we decided to rederive all equations and results in Ref.~\cite{NrFf1976}.

We first rederive the scattering amplitude for $1S-2S$ transition induced by the SEPE process. We 
should mention that authors in Ref.~\cite{NrFf1976} used the following form $g_l(\nu_{i/f},r,r')$ 
in the Schr\"{o}dinger-Coulomb green function of Eq.~(\ref{coulombgreen})~\cite{Am1977},
\begin{align}
 g_{l}(\nu_{i/f},r,r')&=-\frac{2(2\nu_{i/f})^{2l+1}}{[(2l+1)!]^2}e^{-\nu_{i/f}(r+r')}(rr')^l
\sum_{n=l+1}^{\infty}\Big(k-\frac{1}{\nu_{i/f}}\Big)^{-1}\frac{(n+l)!}{(n-l-1)!}\times\nonumber\\
&\times \: _1F_1(l+1-n;2l+2;2\nu_{i/f} r) \: _1F_1(l+1-n;2l+2;2\nu_{i/f} r'),
\label{coulombgreenappe}
\end{align}
where $\:_1F_1 $ refers to the confluent Hypergeometric function. Note that for simplicity, we drop 
indexes of $r_1$ and $r'_1$ in the rest of our calculation. Putting Eq.~(\ref{coulombgreen}) into 
Eq.~(\ref{finalamplitude}) yields the following formula for the scattering amplitude of the $S-S$ 
transition induced by the SEPE process,
\begin{align}
A_{if}&=\frac{E}{(2\pi)^2q^2}\frac{{\bm q}\cdot {\bm \epsilon}}{3}
\int r^3 r'^3 R^{*}_{n_f,0}(r)R_{n_i,0}(r')
\Big[g_{l=1}(\nu_f,r,r')+g_{l=1}(\nu_i,r,r')\Big]dr dr',
\label{radpartamp}
\end{align}
where $R_{n_f,0}(r)$ and $R_{n_i,0}(r')$ are the radial part of the wave function in the ground and excited 
states. Note that $\nu_{i/f}=Z/n_{i/f}$ or $\nu_{i/f}={\cal E}_{i/f} \pm \omega_{\rm L} $. Inserting 
Eq.~(\ref{coulombgreenappe}) into Eq.~(\ref{radpartamp}) results in the following relation for the scattering 
amplitude,
\begin{align}
 A_{if}&=\frac{-E}{(2\pi q)^2}\frac{\bm{q}\cdot 
\bm{\epsilon}}{3}\frac{2^4}{[3!]^2}
\Big[\nu_f^3S(\nu_f)+\nu_i^3S(\nu_i)\Big].
\label{radpartamp1}
\end{align}
In Eq.~(\ref{radpartamp1}), $S(\nu_{i/f})$ reads
\begin{align}
 S(\nu_{i/f})=\sum_{n=2}^{\infty}n(n^2-1)G(n,\nu_{i/f}),
\label{sumformula}
\end{align}
where the summation is over all possible intermediate states. The function $G(n,\nu_{i/f})$ is
\begin{subequations}
\begin{align}
G(n,\nu_{i/f})=(n-\frac{1}{\nu_{i/f}})^{-1}F_{n_f}(n,\nu_{i/f}) F_{n_i}(n,\nu_{i/f})
\label{funtionsa}
\end{align}
\begin{align}
 F_{n_{i/f}}(n,\nu_{i,f})=\int_0^\infty r^{4}e^{-\nu_{i/f}r}R_{n_{i,f}0}(r)\:_1F_1(2-n;4;2\nu_{i/f}r) dr,
\label{funtionsb}
\end{align}
\label{funtions}
\end{subequations}
where $R_{n_{i,f},0}(r)$ is the radial part of the wave function.
 
Since we are interested in calculating the excitation cross section of the $1S-2S$ transition 
induced by the SEPE process based on Eq.~(\ref{totalperpen}) (for ${\bm \epsilon}\perp {\bm k_i}$ 
geometry), it is required to calculate  the integrals of Eq.~(\ref{funtionsa}) for the initial ($n_i=1$) and 
final ($n_f=2$) states. These integrals read, 
\begin{subequations}
\begin{align}
F_{1S}(n,x)&= \frac{4!}{(1+x)^6}\left(\frac{1-x}{1+x}\right)^{m-3}(2-nx)
\label{1sintegral}
\end{align}

\begin{align}
F_{2S}(n,x)&=-\frac{12}{\sqrt{2}}\frac{x^2}{(1/2+x)^8}\left(\frac{1/2-x}{1/2+x}\right)^{n-4}
\left[n^2-(x+\frac{9}{4x})n+(2+\frac{1}{x^2})\right]
\label{2sintegral}
\end{align}
\label{2sintegraltotal}
\end{subequations}
After inserting Eqs.~(\ref{1sintegral}) and (\ref{2sintegral}) into Eq.~(\ref{funtionsa}), one can readily 
calculate $S(\nu_{1S/2S})$ of Eq.~(\ref{sumformula}),
\begin{align}
S(\nu_{1S/2S})&=\frac{288x^2}{\sqrt{2}}\sum_{n=2}^\infty \frac{n(n^2-1)\left[n^2-n(x+9/4)+1/x^2+2\right]}{n-1/x}
\left(\frac{1-x}{1+x}\right)^{n-3}\left(\frac{1/2-x}{1/2+x}\right)^{n-4}\nonumber\\
&\times\frac{2-nx}{\left(1/2+x\right)^8\left(1+x\right)^6}
\label{sumapp}
\end{align}
This equation should be compared with Eq.~($5$) from Ref.~\cite{NrFf1976}, which is
\begin{align}
S'(\nu_{1S/2S})&=\frac{256x^2}{3\sqrt{2}}\sum_{n=2}^\infty \frac{n(n^2-1)\left[n^2-n(x+9/4)+1/x^2+2\right]}{n-1/x}
\left(\frac{1-x}{1+x}\right)^{n-3}\left(\frac{1/2-x}{1/2+x}\right)^{n-4}\nonumber\\
&\times\frac{1}{\left(1/2+x\right)^8\left(1+x\right)^6}
\label{sumappb}
\end{align}  
Clearly, one can observe inconsistencies between Eq.~(\ref{sumapp}) and Eq.~(\ref{sumappb}) from Ref.~\cite{NrFf1976}.
There is one more reason for the deviation between our numerical result and available literature sources, which is related to Eq.~(\ref{anugualrperpen}). As we discussed in the text, the authors in Refs.~\cite{NrFf1976,NrFf1978} employed a wrong expression for $A_{\perp}$, see Eqs.~(\ref{anugualrperpen}) and (\ref{anugualrperpenref})~[see also Eq.~($4a$) from Ref.~\cite{NrFf1976} and Eq.~(11) from Ref.~\cite{NrFf1978}].
These deviations are responsible for inconsistencies between our numerical result of the excitation cross section of 
$1S-2S$ transition and the one in Ref.~\cite{NrFf1976}.

Performing summation in Eq.(\ref{sumapp}) and inserting it into Eqs.~(\ref{radpartamp1}) and (\ref{totalperpen}) 
yield the total cross section $0.266$ in units of $\pi a_0^2$ for the $1S-2S$ transition induced by the SEPE 
process in Hydrogen atom, which can be compared with the $1S-2S$ result in Table~\ref{tab1}. It should be 
pointed out that this calculation is done for the (electron) projectile with an initial energy $100~{\rm eV}$. We 
add that the excitation cross section of the $1S-2S$ transition obtained in Ref.~\cite{NrFf1976} is $1.83$ in units 
of $\pi a_0^2$ for the (electron) projectile with energy $100~{\rm eV}$.  

\end{document}